\documentclass[pre,superscriptaddress,showpacs,nofootinbib,floatfix,preprint]{revtex4}
\usepackage{amsmath}
\usepackage{amssymb}
\usepackage{epsfig}

\usepackage{dcolumn}
\usepackage{bm}     

\begin{document}

\preprint{PREPRINT}

\date{\today}

\title{Incorporation of excluded volume correlations into Poisson-Boltzmann theory}

\author{Dmytro Antypov}\email{antypov@mpip-mainz.mpg.de}\affiliation{Max-Planck-Institut
  f\"ur Polymerforschung, Ackermannweg 10, 55128 Mainz, Germany}
\author{Marcia C.\ Barbosa}\email{barbosa@if.ufrgs.br}\affiliation{Instituto de
  F\'{\i}sica, UFRGS, 91501-970, Porto Alegre, RS, Brazil}
\author{Christian
  Holm}\email{c.holm@fias.uni-frankfurt.de}\affiliation{Frankfurt Institute for
  Advanced Studies (FIAS), Johann Wolfgang Goethe - Universit\"at, Max-von-Laue-Str{\ss}e 1,
D-60438  Frankfurt am Main, Germany}
\affiliation{Max-Planck-Institut f\"ur  Polymerforschung, Ackermannweg 10, 55128 Mainz, Germany}

\begin{abstract}
  We investigate the effect of excluded volume interactions on the electrolyte
  distribution around a charged macroion. First, we introduce a criterion for
  determining when hard-core effects should be taken into account beyond
  standard mean field Poisson-Boltzmann~(PB) theory. Next, we demonstrate that
  several commonly proposed local density functional approaches for excluded
  volume interactions cannot be used for this purpose. Instead, we employ a
  non-local excess free energy by using a simple constant weight approach.  We
  compare the ion distribution and osmotic pressure predicted by this theory
  with Monte Carlo simulations. They agree very well for weakly developed
  correlations and give the correct layering effect for stronger ones.  In all
  investigated cases our simple weighted density theory yields more realistic
  results than the standard PB approach, whereas all local density theories do
  not improve on the PB density profiles but on the contrary, deviate even more
  from the simulation results.
  
\end{abstract}


\maketitle

\section{Introduction}
\label{introductions}
Understanding the behavior of charged macroions in solution is an important
problem in fundamental science \cite{barrat96a} as well as in industrial
\cite{hara93a} and biological applications \cite{holm01a}.  Charged stabilized
colloidal dispersions are present in paints, inks, pharmaceutical products and
are used in the fabrication of nanostructured
materials~\cite{gast98a,frenkel02a,velev00a}. These systems serve also as a
primitive model for the crowded cellular environment that represents numerous
biomacromolecules and cellular polymers~\cite{ellis03a,minton95a}.  What all
the applications above have in common is that when a charged macroion is
immersed in an electrolyte solution, it is surrounded by counterions to
balance the surface charge. The charged macroion surface along with the
neutralizing diffuse layer of counterions is usually refereed as the electric
double layer, the understanding of which is crucial for describing the
behavior of such systems.  For instance, the stability of colloidal dispersion
depends on the distribution of small ions around the colloid.  The
electrophoretic mobility of the solution also can be rationalized in terms of
the ion distribution~\cite{gonzalez-tovar85a,lozada01a,tanaka02a,lobaskin04b}
and most of the electrochemical reactions occur in this interfacial
region~\cite{yang02a}.

As a result, there has been a considerable effort to describe the density
profile around the macroion for different macroion geometries.  The earliest
theory that had significant success was the Poisson-Boltzmann~(PB) approach.
Its versions for planar geometry, the so called Gouy-Chapman
theory~\cite{gouy10a,chapman13a}, can be solved exactly. It also has an
analytical solution for an infinitely long linear macroion confined to a
cylindrical cell \cite{fuoss51a,alfrey51a}, whereas only a numerical solution
can be obtained in the case of a spherical geometry.  The major flaw of this
mean-field approach is that it neglects all correlations between the ions.
For a long time, integral equation theories have been developed to adequately
describe dense systems of electrolytes,
and recently field theories
have become very popular in calculating correlation corrections to the mean
field PB approach, see i.e. Refs. ~\cite{holm01a,grosberg02a,levin02a} for
overviews. However, since the treatment of size effects is mixed with the
electrostatic correlations, in many approaches it becomes difficult to
identify the role of each effect. And finally integral equation theories work
well at high densities when excluded volume contributions are very strong,
whereas they are problematic in the low density regime.

It would therefore be desirable to have a theoretical framework which retains
the simplicity of the early attempts, but also accommodates correlation
effects -- something that can be done within density functional theories.  It
is possible to rigorously rewrite the partition function of, say, a system of
charged colloids, as a density functional~\cite{loewen93a}, in which the
contribution beyond mean field is included as an additive correlation
correction to the free energy density. The functional form of this correction
is unknown and one has to use a reasonable Ansatz for it.  The spirit is very
similar to the fundamental problem of integral equations, where one also has
to make an educated guess (namely, the closure relation). However, in the case
of a functional this involves a free energy density expression rather than a
relation between two- and three-point functions.  It thus relies on a
different kind of intuition and thus permits some complementary insight.

A number of density functional prescriptions for taking both size and
electrostatic correlations into account have been
proposed~\cite{patra93a,gillespie03a,yu04a,wu04a}.  These theories are able to
reproduce to some extent the density profile of charged systems. However,
since they treat both size and electrostatic correlations together, the origin
of the result is not clear. Recently we adopted a different approach. We
studied systems of point-like counterions (therefore no size effects) and
addressed the question of when the electrostatic correlations become relevant.
For treating these correlations we proposed the Debye-H\"uckel-Hole-Cavity
functional~\cite{barbosa00a,barbosa04a}. This theory relies on a
Debye-H\"uckel treatment of the One Component Plasma
(OCP)~\cite{salpeter58a,abe59a,baus80a}, in which the short-distance failure
of linearization is overcome by postulating a correlation hole. Since beyond a
certain density the resulting OCP free energy density is a concave function of
density, this favors the development of inhomogeneities.  In the pure OCP
these are balanced by the homogeneously charged background. However, if one
uses the OCP free energy density as a correlation correction to the mean-field
functional describing the double layer at a charged surface, one has all the
charge opposite to the counterions located on that surface, rather than
homogeneously distributed as a stabilizing background. The consequence is that
the double layer becomes unstable and all ions collapse onto the surface, an
effect which has been termed ``structural
catastrophe''~\cite{groot91a,penfold90a}.  To prevent this effect we
introduced a spherical exclusion region where no background can be found. The
prescription for finding the size of such an exclusion serves both to keep the
theory self-consistent and to establish the range of validity of the PB
approach. Comparisons of the ionic charge distribution around a charged cylinder
and a charged sphere showed a very good agreement with the simulations for
both monovalent and trivalent counterions~\cite{barbosa04a}.

Having studied how to take into account the electrostatic correlations, we
address in this paper the relevance of excluded volume correlations. We
present a validity criterion for the PB approach by constructing a parameter
whose value approximately indicates when excluded volume correlations are
expected to become relevant. We then test several local density approaches,
that have been advocated~\cite{borukhov97a,borukhov04a}, demonstrating that
they all fail to take the size correlations into account, and that they even
lead to an instability of the solution beyond a certain ion size. In order to
circumvent this, a weighted density functional based on a simplified Tarazona
approach~\cite{tarazona84b,tarazona85a,tarazona87a} is introduced. Our results
are compared with Monte Carlo (MC) simulations, showing very good agreement
for the cases of moderately developed hard core correlations, and even for
strongly electrostatically interacting systems in both zero salt and non-zero
salt cases.

The remainder of the paper is organized as follows. In Sec. II we
discuss validity of the PB approach for a colloidal system with
non-point like counterions and show how the size effect can be
incorporated into the model. Details of the used numerical methods
are given in Sec. III. These are followed by the results and
discussions presented in Sec. IV, and our conclusion in Sec. V.

\section{Size Correlations within Density Functional Theory}
\label{DFT}

\begin{figure}[!ht]
  \begin{center}
  \includegraphics[height=8.6cm]{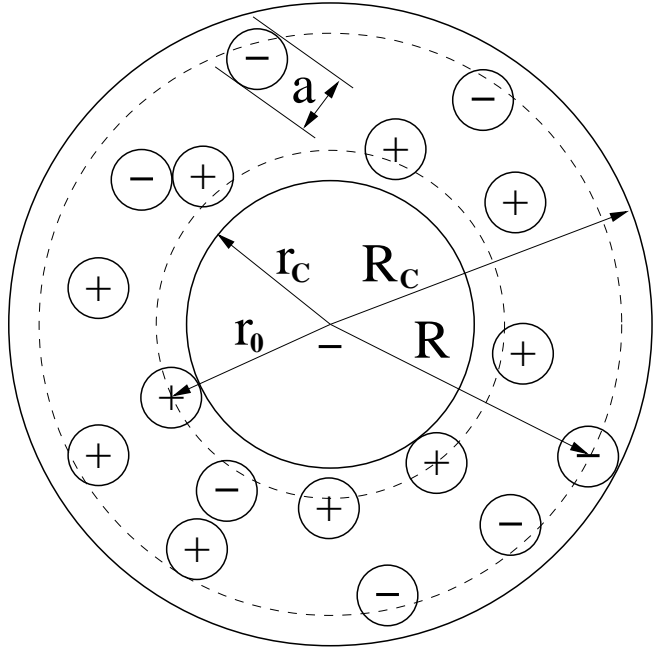}
  \caption{Colloidal cell model. The system is defined by five independent parameters:
   number of counterions, $N$, number of salt ions, $N_s$, and three
   characteristic sizes: $r_0$, $R$ and $a$ measured in units of $l_B v^2$.
   See Appendix for more details.}
  \label{cell}
  \end{center}
  \end{figure}

Consider a spherical colloid of radius $r_c$ and negative charge
$Z$, which is located in the center of a spherical cell of radius
$R_c$. This cell represents a bulk colloidal solution with colloid
volume fraction $\phi=(r_c/R_c)^3$. The counterions are taken as
positively charged hard spheres of diameter $a$ and valence $v$
and $N = Z/v$ of them provide the neutrality of the cell.
The solvent is modeled as uniform dielectric background of dielectric constant
$\epsilon$, and the strength of the electrostatic interactions is
defined by Bjerrum length
\begin{eqnarray}
 \l_B = \frac {q^2} {4 \pi \epsilon_0 \epsilon k_B T} \;,
\end{eqnarray}
where $q$ is the unit charge. In the non-zero salt case, $N_s$
positive and $N_s$ negative salt ions are also included. Here we
assume that all ions have the same size and valence as the
counterions. The average charge distribution is described by local
densities $n_-(r)$ for the coions and $n_+(r)$ for the counterions
and positive salt ions. These are defined for $r_0 \leq r \leq R$,
where $r_0=r_c+a/2$ is the distance of closest approach between
the macroion and the particles and $R=R_c-a/2$ (See
Fig.~\ref{cell}).
%
%
%
%
Therefore, the volume fraction of the bulk electrolyte $\phi_e$ is given by
\begin{eqnarray}
  \label{phi_e}
 \phi_e = \frac {a^3(N + 2N_s)} {8(R_c^3-r_c^3)} \; .
\end{eqnarray}
The effective surface charge density $\sigma$ should be
defined in terms of $r_0$ rather than $r_c$, i.e. $\sigma = Z /(4\pi r_0^2)$.
This will be used later when defining the plasma
parameter of the system.

The central task of a density-functional theory is to derive an
analytical expression for the free Helmholtz energy functional
that upon minimization gives the density profiles of the free ions
in solution. Its simplest form is given by the Poisson-Boltzmann
functional, namely
\begin{eqnarray}
  \label{FPB}
  F_{PB}&=&
  \int_{}^{} d^3r \;
   \Big \{ k_B T \, n_+({\bf r}) \big[ \ln \big( n_+({\bf r}) \lambda^3 \big) -1 \big]\,
+  k_B T\, n_-({\bf r}) \big[ \ln \big( n_-({\bf r}) \lambda^3 \big) -1
\big]\, + \, f_{el}({\bf r}) \Big\} \; ,
\end{eqnarray}
which includes the translational entropy of the ions ($\lambda$ is
the thermal de Broglie wavelength) and all electrostatic
interactions represented by $f_{el}({\bf r})$. The electrostatic
interactions within the mean-field approximation are given by
\begin{eqnarray}
  \label{fel}
f_{el}({\bf r}) &=& v q [n_+({\bf r})-n_-({\bf r})]\, \psi({\bf r}) \;,
\end{eqnarray}
where $\psi({\bf r})$ is the total electrostatic potential created at
position $({\bf r})$ by the fixed macroion and all ions together. The
minimization of Eq.~(\ref{FPB}) with respect to $n_{+}({\bf r})$ and
$n_{-}({\bf r})$ gives the Boltzmann density distributions
\begin{equation}
\label{npb} n_{\pm}({\bf r}) = n^0_{\pm} e^{\mp\beta v q \psi({\bf r})}\;,
\end{equation}
where parameters $n^0_{+}$ and $n^0_{-}$ are defined by the charge neutrality
condition. In spherical geometry, Eq.~\ref{npb} together with the Poisson equation
\begin{equation}
\nabla^2 \psi(r)=-\frac{4qv\pi}{\epsilon}[n_+(r)-n_-(r)] \;
\end{equation}
and the boundary conditions at $r=r_0$ and $r=R$ comprises a fully defined
Poisson-Boltzmann problem.  The problem with this approach is that the ions
are considered as point charges in some average electric field and both
electrostatic and excluded volume correlations between them are not taken into
account.  When do these correlations matter? Comparing the PB predictions to
simulation results, one has found out, that, for point-like ions,
electrostatic correlations become relevant when the plasma parameter,
$\Gamma_{2d}=\sqrt{\pi\sigma l_B^2v^3}$, becomes larger than 1
~\cite{rouzina96a,barbosa00a,joensson01a,moreira02a,deserno03a,barbosa04a}.
Here we address the question under what condition the excluded volume effects
become significant. For a uniform hard sphere liquid we know that the radial
distribution function changes from monotonically decaying to non-monotonic at
volume fractions of the order of $\phi \approx 0.2$, which will be the
reference volume fraction in our further estimates.  In the case of confined
liquids, ions tend to concentrate close to the surface and their concentration
at the surface, especially in charged systems, may be much higher than that in
the bulk~\cite{messina02d}. Therefore, to access the hard-core effects one
should consider the average volume fraction within the first layer of
counterions close to the colloid surface.  Since the colloid is much larger
than the ions it can be approximated as (for simplicity the salt is not
included)
\begin{equation}
\label{phi1} 
\phi_s=\frac{\pi a^2}{6}\int_{0}^{a} n_{PB}(x) dx \;,
\end{equation}
where \mbox {$n_{PB}(x)\ell^3= \Gamma_{2d}^4/\pi(2\bar{x}\Gamma_{2d}^2+1)^2$}
is the exact solution obtained in the planar
geometry~\cite{gouy10a,chapman13a,wennerstroem82a}.  Here $\bar{x}=x/\ell$,
where $\ell=l_{B} v^2$.  After integration we find that the volume fraction
close to the macroion is given by
\begin{equation}
\label{phi2}
\phi_s=\frac{\hat{a}^3\Gamma_{2d}^4}{3(2\Gamma_{2d}^2\hat{a}+1)}
\;,
\end{equation}
where $\hat{a}\equiv a/\ell$.  Then, for $\phi_s \lesssim 0.2$, there are only
weakly developed excluded volume correlations, and the PB approximation should
be still valid. For larger values of $\phi_s$ we expect to see some layering effects
close to surfaces. This criterion is strictly valid only for a planar
geometry, but is expected to approximately hold for sufficiently large
spherical or cylindrical macroions where the curvature effects ($\propto 1/R$)
are negligible. An analogous formula which takes the curvature into account
can also be derived for a cylindrical PB cell for which
the contact density is known analytically. 
Since electrostatic correlations were not taken into account, we do
not expect this simple analysis to hold for $\Gamma_{2d} \gtrsim 2$.  Beyond this
value, the force-distance curves between charged plates cease to be monotonic,
and beyond $\Gamma_{2d}\simeq 2.45$ attractions even between like charged
macroions can occur~\cite{moreira02a}.  These effects are results of
correlations between different double layers (like, for instance, ion
interlocking \cite{jensen97a,deserno03a}) that are stronger than the size
effects we are describing here. Note that, Eq.~(\ref{phi2}) is designed to
give a limit of validity of the mean-field approach.  For high ionic radius
$\phi_s$ can become larger than $1$, loosing its relation to the actual volume
fraction in the system.

Correlations can be included into the PB model by adding to the PB free energy
$F_{PB}$ an excess free energy term,
\begin{equation}
\label{f} F=F_{PB}+F_{ex} \; .
\end{equation}
%
The excess free energy $F_{ex}$ originates from internal interactions within
the system and it is unknown. In principle, it can account for both the
hard-core repulsion and the electrostatic correlations.
Since we want to test the excluded volume effects within the range of ionic
strength in which they overcome the electrostatic correlations, the latter
will be neglected within our approach.

There are a number of various functionals which can be used to include
hard-core effects in uniform liquids of a given density $n$. Within the local
density approximation (LDA) one can choose one of these free energy density
expressions, $f_{ex}[n]$, and replace the uniform density by a local one so
that the total excess free energy reads
\begin{eqnarray}
  \label{fex}
F_{ex} &=& k_B T \int d^3r \, n(r) f_{ex}[n(r)] \;.
\end{eqnarray}
For our system we can take $n(r) = n_+(r) + n_-(r)$ which means that the
hard-core effects are treated identically for both positive and negative ions.
The idea behind Eq.~(\ref{fex}) is quite simple. A particle at ${\bf{r}}$ is
supposed to be affected by only the particles around it, in a range given by
the interaction. If the range of the interparticle interaction is much smaller
than the typical length for variations in $n(r)$, the system can be divided in
small subvolumes of nearly constant density and each of them can be
treated as part of a homogeneous system. If we take ions as charged hard
spheres, we can use the free energy density derived, for example, from the
Carnahan-Starling (CS) equation of state~\cite{carnahan69a},
namely
\begin{eqnarray}
  \label{fcs}
f_{CS}[\phi(r)]&=& \frac{ \phi(r)(4-3\phi(r))}{(1-\phi(r))^{2}}
\end{eqnarray}
where $\phi(r)=\pi\, a^3 n(r)/6$ is the volume fraction occupied
by the free ions. For denser liquids, the accuracy might be
improved by using the more precise virial expansion for the Percus-Yevick
theory for hard spheres~\cite{mcquarrie76a}, namely
\begin{eqnarray}
  \label{fvir}
f_{vir}[\phi(r)]&=&\Big\{ 4
\phi(r)+5\phi(r)^2+6.12\phi(r)^3+7.02\phi(r)^4+7.905\phi(r)^5+9.4208
\phi(r)^6 \Big \} \; .
\end{eqnarray}
The last two expressions agree up to second order in local volume
fraction $\phi(r)$.

A simple form of $F_{ex}$ can also be
derived from the free volume (fv) expression for a lattice
gas~\cite{borukhov97a,borukhov04a}, namely
\begin{eqnarray}
  \label{ffv1}
F_{fv1} &=& k_B T \int_{}^{} d^3r \, \left[\frac{1}{a^3}-n(r)\right]\,
\ln \left( 1-n(r)a^3) \right) \;,
\end{eqnarray}
where $a$ is the lattice spacing.
With this form of functional the excluded volume effects can be
explicitly incorporated into the PB equation. However, its
density expansion is different from
that of hard sphere. Another expression based on the free volume
concept which can be found in Ref.~\cite{hansen90a}
\begin{eqnarray}
  \label{ffv2}
F_{fv2} &=& - k_B T \int_{}^{} d^3r \, n(r)
\ln \left( 1-\frac{\phi(r)}{2} \right)\
\end{eqnarray}
gives lower order density terms similar to Eq.~(\ref{fcs}) and
Eq.~(\ref{fvir}).

The assumption of smooth variations of $n(r)$ being within the
characteristic range of hard-core interactions is only valid for
sufficiently small ionic radii $a$. Consequently, as we will show later, all the
functionals above underestimate the densities close to the colloid
and completely fail to give the correct density profile in the
more interesting cases where stronger variations in $n(r)$ are
observed. Moreover, all expressions above have a singularity at
certain value of volume fraction. This reflects the fact that the bulk density 
cannot exceed some upper limit. If a local density is higher than that,
for example due to a charged surface, the local density functional does not
converge and no density profile can be obtained.

In order to circumvent this difficulty, a number of weighted
density approaches (WDAs) have been
developed~\cite{tarazona84b,tarazona85b,tarazona85a,tarazona87a,rosenfeld89a,
rosenfeld93a,nordholm80a} for the description of neutral hard
sphere solutions, and some rather complex methods have already
been proposed for charged suspensions
\cite{patra93a,gillespie03a,yu02a}. The prescription we have
followed here, in the spirit of the generalized van der Waals
theory of Nordholm and co-workers~\cite{nordholm80a}, is to
introduce the non-locality of the free-energy density functional
through a coarse-grained density distribution $\bar{n}(r)$, which at
each point is a non-local functional of the local density $n(r)$.
This can be pictured as a mean density around point ${\bf r}$
averaged over a volume related to the range of the interactions.
  In this context, the local density in Eq.~(\ref{fex})
is replaced by some weighted density $\bar{n}(r)$, namely
\begin{eqnarray}
  \label{fexnew}
 F_{ex} &=& k_B T \int_{}^{} d^3r \, n(r) f_{ex}[\bar{n}(r)] \;
\end{eqnarray}
where
\begin{eqnarray}
\label{barn} \bar{n}(r)=\int_{}^{} d^3r' w(|{\bf r}-{\bf r}'|)
n(|{\bf r}'|)\; .
\end{eqnarray}
The weight function $w(|{\bf r}-{\bf r}'|)$ should be chosen to
give reasonable direct correlation functions which are functional
derivatives of $F_{ex}[\bar{n}]$. The most important are the
first- and second-order correlation functions, defined as
\begin{eqnarray}
\label{corr}
c^{(1)}(r)&=&\frac{\delta F_{ex}[\bar{n}]}{\delta n(r)}\\
c^{(2)}(r,r')&=&\frac{\delta^2 F_{ex}[\bar{n}]}{\delta n(r)\delta
n(r')} \nonumber \; .
\end{eqnarray}

In Tarazona's~\cite{tarazona84b,tarazona85b,tarazona87a} approach one assumes that the
weight itself is also density dependent and can be expanded in powers of
the weighted density as follows
\begin{eqnarray}
\label{tarweight} w(|{\bf r}-{\bf
r}'|)=w_0(r)+w_1(r)\bar{n}(r)+w_2(r)\bar{n}(r)^2+... \; .
\end{eqnarray}
If we substitute this expression into the direct correlation
function in Eq.~(\ref{corr}), the resulting expansion can be set
equal to the direct correlation function of a uniform hard sphere
fluid~\cite{wertheim63a}. This way it is possible to obtain the
weight function that up to second order is given
by~\cite{tarazona84b,tarazona85b,tarazona87a}

\begin{eqnarray}
\label{wtarazona}
w_0(r)&=&\frac{3}{4\pi a^3}\Theta[a-r] \\
\label{wtarazona1}
w_1(r)&=&[0.475-0.648\frac{r}{a}+0.113\left(\frac{r}{a}\right)^2],\;\; r<a \\
&=&[0.288\frac{a}{r}-0.924+0.764\frac{r}{a}-0.187\left(\frac{r}{a}\right)^2],\;\; a<r<2a \nonumber \\
&=&0, \;\; r>2a \nonumber \\
\label{wtarazona2}
w_2(r)&=&\frac{5\pi a^3}{144}[6-12\frac{r}{a}+5\left(\frac{r}{a}\right)^2],\;\; r<a \\
&=&0 \;\; r>a.\nonumber
\end{eqnarray}
Since our aim is not to precisely describe the hard-sphere effects but just to
access their relevance, we will employ the simplest form of the weight
function that is the first term in Eq.~\ref{tarweight} or a constant
weight~\cite{nordholm80a} given by Eq.~\ref{wtarazona}. We will refer to this
weight as~WDA0 whereas the weight defined by
Eqs.~\ref{wtarazona}-\ref{wtarazona2} will be called~WDA2 and will be used to
validate our results.  For a pure hard sphere fluid, WDA0 reproduces the
discontinuity in the direct correlation function predicted by Percus and
Yevick~\cite{percus58a,hansen90a}.  However, it overestimates the range of the
correlation function, especially at high densities, when compared to the
density dependent weights~\cite{tarazona85a,patra93a} or to direction
dependent weights~\cite{rosenfeld93a,gillespie03a,yu02a}.  Having this in
mind, we will concentrate on the systems for which size plays a relevant role
but the differences between the constant weight and more sophisticated
approaches do not affect our main conclusions.

Measuring all lengths in units of $\ell=l_{B} v^2$ reveals that the full
partition function of our cell model depends on five system parameters. The
observables, for example, reduced density profile $\hat{n}(r)=n(r)\ell^3$,
remain constant under rescaling which does not change the following
quantities: the number of counterions $N=Z/v$, the number of salt ions $N_s$,
the reduced distance of closest approach $\hat r_0 = r_0/\ell$, the $r_0/R$
ratio, and the reduced ion diameter $\hat a = a/\ell$ (see Appendix).  The
same holds for PB theory and is also true for our WDA theory. For the WDA it
implies that both the WDA free energy correction and the weight function have
to obey this restriction.
\section{Numerical methods}
\label{montecarlo}

In this Section we give details of the numerical methods used to study the
cell system described in Sec. II. Three different ways were employed to find
the ion distribution in the cell. The first one was a direct Monte Carlo
simulation of the cell model which gave us reference data to test the
theoretical results. The density profile minimizing a given free energy
functional was obtained using numerical iteration until it converged to the
equilibrium charge distribution. Another way of minimizing the functional was
by Monte Carlo sampling. Some technical details of these three methods are
summarized below.

\subsection{Monte Carlo simulation.}
Within this approach we simulate the cell model exactly as it is -- all ions
are taken as charged hard spheres of diameter $a$ confined between two
spherical shells of radii $r_c$ and $R_c$. To gather the statistics of charge
distribution the ions are moved around the cell and a single ion move is
either accepted or rejected according to the usual Metropolis probability:
\begin{equation}
\pi = \min \{1, \exp \left( - \beta \Delta E \right) \} \;,
\end{equation}
where $\Delta E$ is the difference between the system energy after
and before the move. Since the density profile is highly
anisotropic, a combination of two types of moves was found to
improve the efficiency of the sampling. An ion was either inserted
at a random position in the cell or randomly displaced within a
cube centered at its current position. The former allowed for
efficient exploration of low density regions, whereas the latter
proved to be efficient close to the colloid where a successful
insertion of an ion could be a rare event due to the high packing
fraction. The frequency of using one or the other move as well as
the displacement range were adjusted to give an about $50\%$
acceptance rate.

\subsection{Iterative functional minimization.}
When minimizing a functional containing a non-zero correlation
term, Eq.~\ref{npb} becomes dependent on the excess chemical
potential
\begin{equation}
\mu^{ex}_{\pm} (r) = \frac {\delta F_{ex}[n(r)]} {\delta n_{\pm}(r)}
\end{equation}
and reads
\begin{equation}
\label{npbex} n_{\pm}(r) = n^0_{\pm} e^{ \mp \beta v q \psi(r) - \beta \mu^{ex}_{\pm}
(r)} \; .
\end{equation}
Once the expressions for $\mu^{ex}_{\pm}$ are derived for a given
functional, we can find the ion distribution which satisfies both
Poisson equation and Eq.~\ref{npbex}. Integrating the Poisson
equation over a spherical shell of radius $r$, and using the Gauss
theorem, an integro-differential equation for the electric field
$E(r)$ can be obtained. Consequently, the optimum density profile
can be obtained from the numerical iteration of this equation
until convergence is achieved.

\subsection{Monte Carlo functional minimization.}
Within this approach the ion position is described only by its
distance from the colloid, $r$, and this uniquely defines the
density distribution, $n(r)$. Each MC step consists of moving an
ion to a new trial position $r_0 < r < R$. This move is either
accepted or rejected with probability~\cite{deserno00d}
\begin{equation}
\pi = \min \{1, \exp \left( - \beta \Delta F \right) \} \;,
\end{equation}
where $\Delta F$ is the free energy difference after and before
the move and it is explicitly given by the functional we minimize.
This method was found to be more stable and worked much faster
than the iterative procedure, though the final result did not
depend on the numerical approach used.

\section{Results and discussion}
\label{results}

In this Section we compare how well the different density
functional approaches described in Sec. II capture the excluded
volume interactions. First, we apply numerical techniques
described in Sec. III to two colloidal systems already studied in
the literature~\cite{borukhov04a}. Then we perform a systematic
analysis of hard-core effects by investigating a number of
different systems with and without added salt.

\begin{figure}[t]
  \begin{center}
    \includegraphics[height=8.0cm,angle=-90]{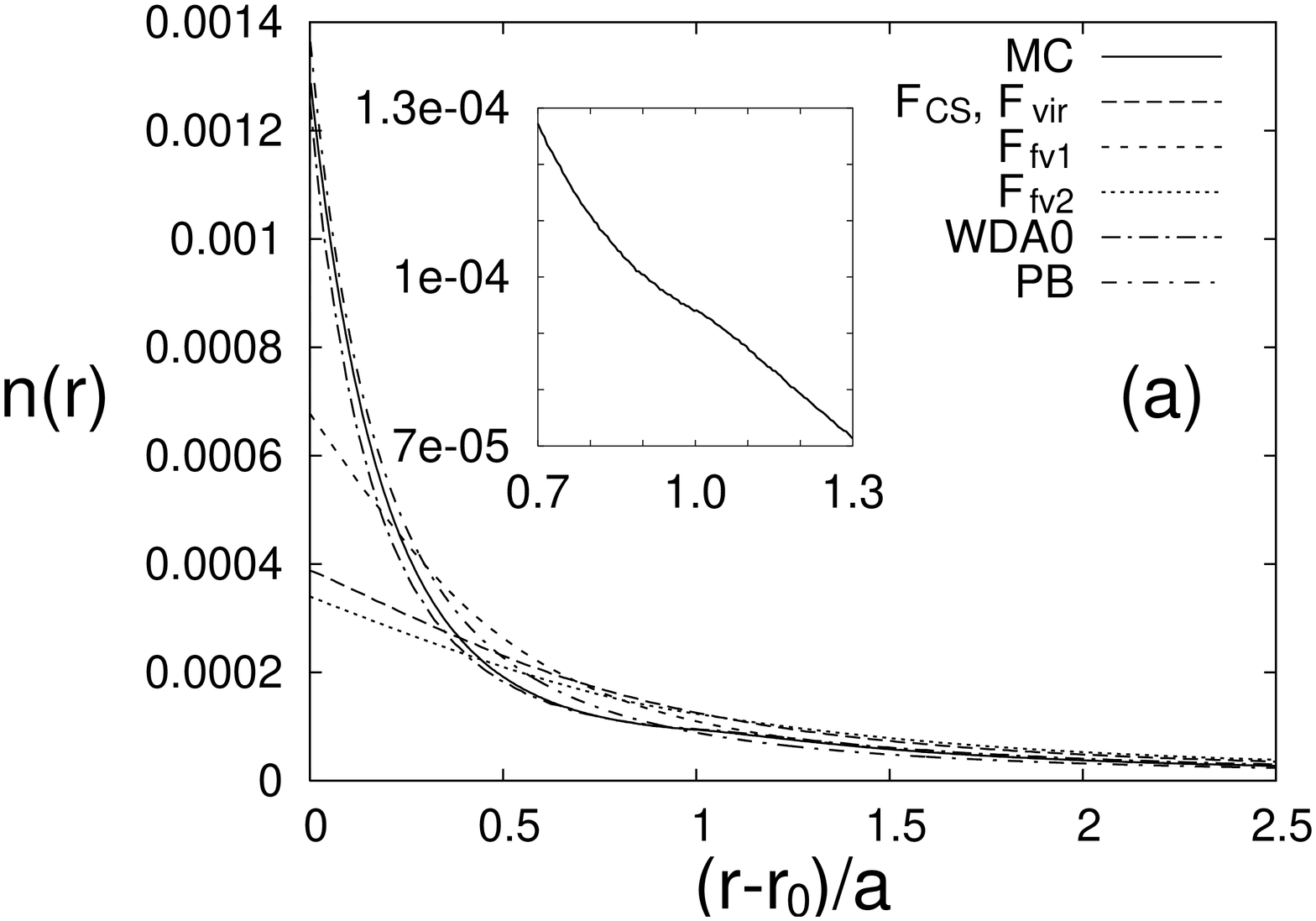}
  \includegraphics[height=8.0cm,angle=-90]{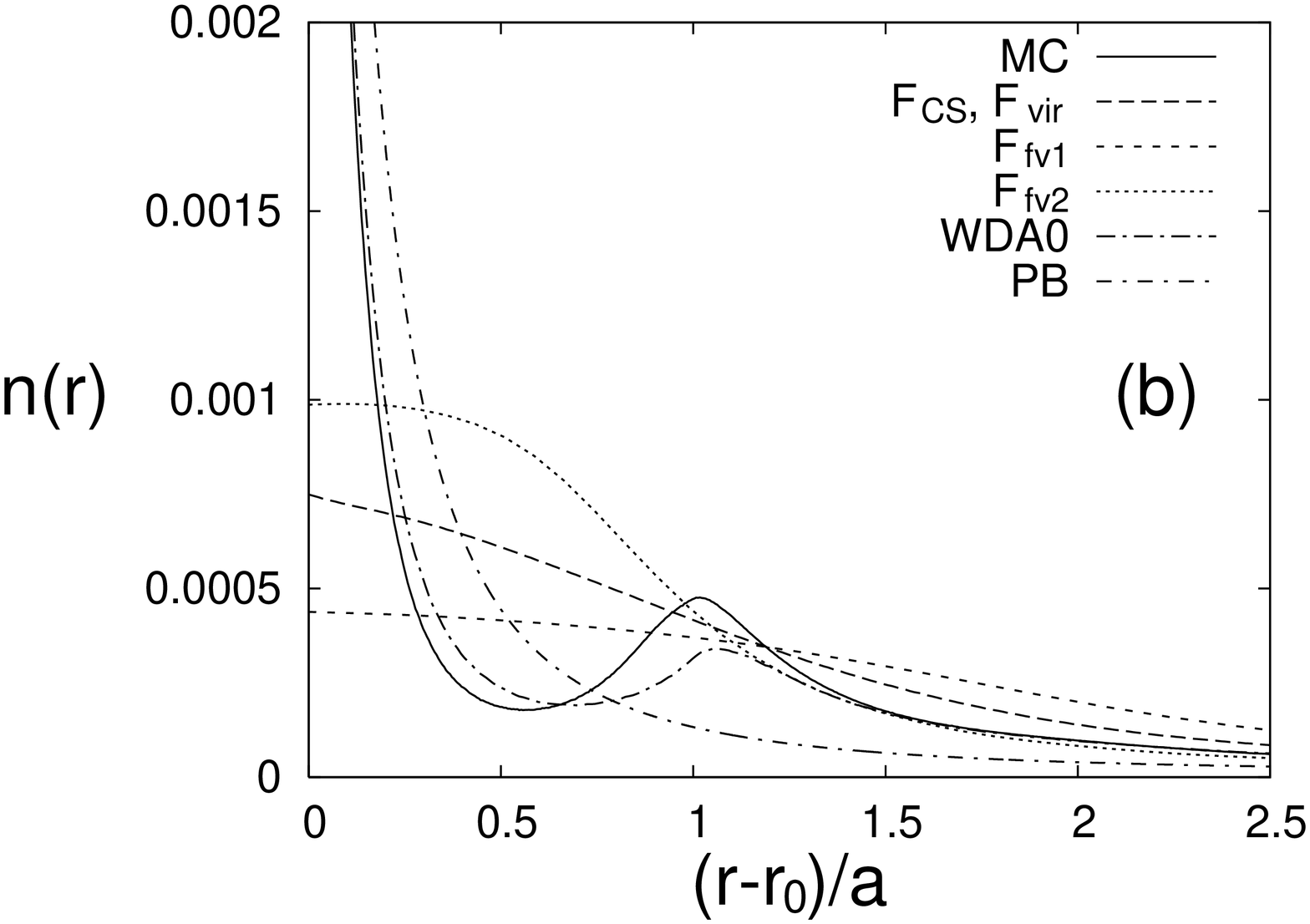}
  \caption{Ion distribution close to the colloid surface measured in systems with
   $r_0 = 50$ \AA, $R = 100$\AA, $a=10$\AA, $l_B = 7$\AA,
   and containing (a) $N=200$
   and (b) $N=500$ monovalent ions. Each curve corresponds to a particular method used: MC
   is the result of the Monte Carlo simulation; $F_{CS}$, $F_{vir}$, $F_{fv1}$ and $F_{fv2}$
   are obtained using
   LDA with a form of the excess free energy given by Eqs.~\ref{fcs}-\ref{ffv2},
   correspondingly; WDA0 is the constant weight curve
   and PB is the numerical solution of PB equation without any hard-core corrections.}
  \label{bor}
  \end{center}
\end{figure}

We start from considering two salt free systems which were also
used in~\cite{borukhov04a} to study the excluded volume effect in
colloidal solutions. In both systems $r_0 = 50$\AA, $R = 100$\AA,
$a=10$\AA, and $l_b = 7$\AA. The number of monovalent ions is
different and it is $N=200$ in system~(a) and $N=500$ in
system~(b). Note that already for 200 ions some packing effects
are expected to be seen because $\phi_s=0.25$. Figure~\ref{bor}
shows the ion density distribution close to the colloid
%
%
obtained using different approaches for both (a) and (b).  Comparison between
the PB and the MC curves for system~(a) reveals that hard-core repulsion
decreases condensation by pushing ions away from the colloid. This effect is
captured well by the WDA, whereas it is significantly overestimated by all
LDAs -- the predicted contact densities are too low. The highest packing
fraction achieved in LDA calculations for system~(a) was below the critical
value and the equations converged.

In the case of 500 ions, the LDA with functional given by
Eq.~\ref{ffv1} was found to be numerically unstable. The iterative
functional minimization failed to converge, while some convergence could
be still achieved by explicitly limiting the highest density at
$a^{-3}$ in the MC sampling of the functional. The result, however, 
depended on the number of bins and therefore was not physical. A
plateau close to the colloid, as seen in~\cite{borukhov04a}, was
observed only for relatively large bin sizes, while smaller bins
resulted in a saw-like density profile (not shown here). The other
LDAs, while still converging, overestimated hard-core effects and
also resulted in unphysical profiles. Moreover, all local
approaches missed the layering clearly captured by WDA0 and 
observed in simulations at distance $a$ from the colloidal surface (Fig.~\ref{bor}(b)).

\begin{figure}[!ht]
  \begin{center}
  \includegraphics[height=8.0cm,angle=-90]{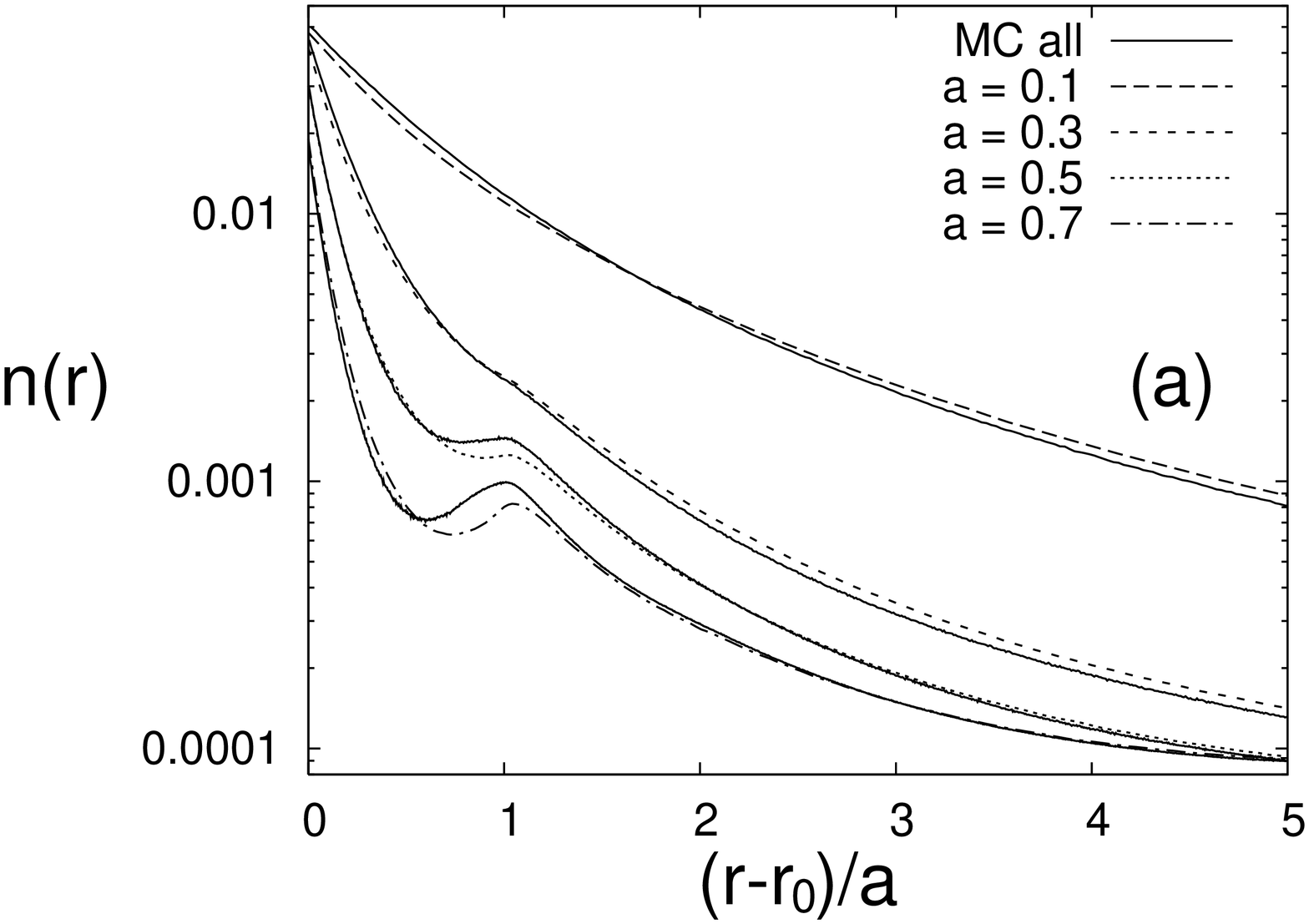}
  \includegraphics[height=8.0cm,angle=-90]{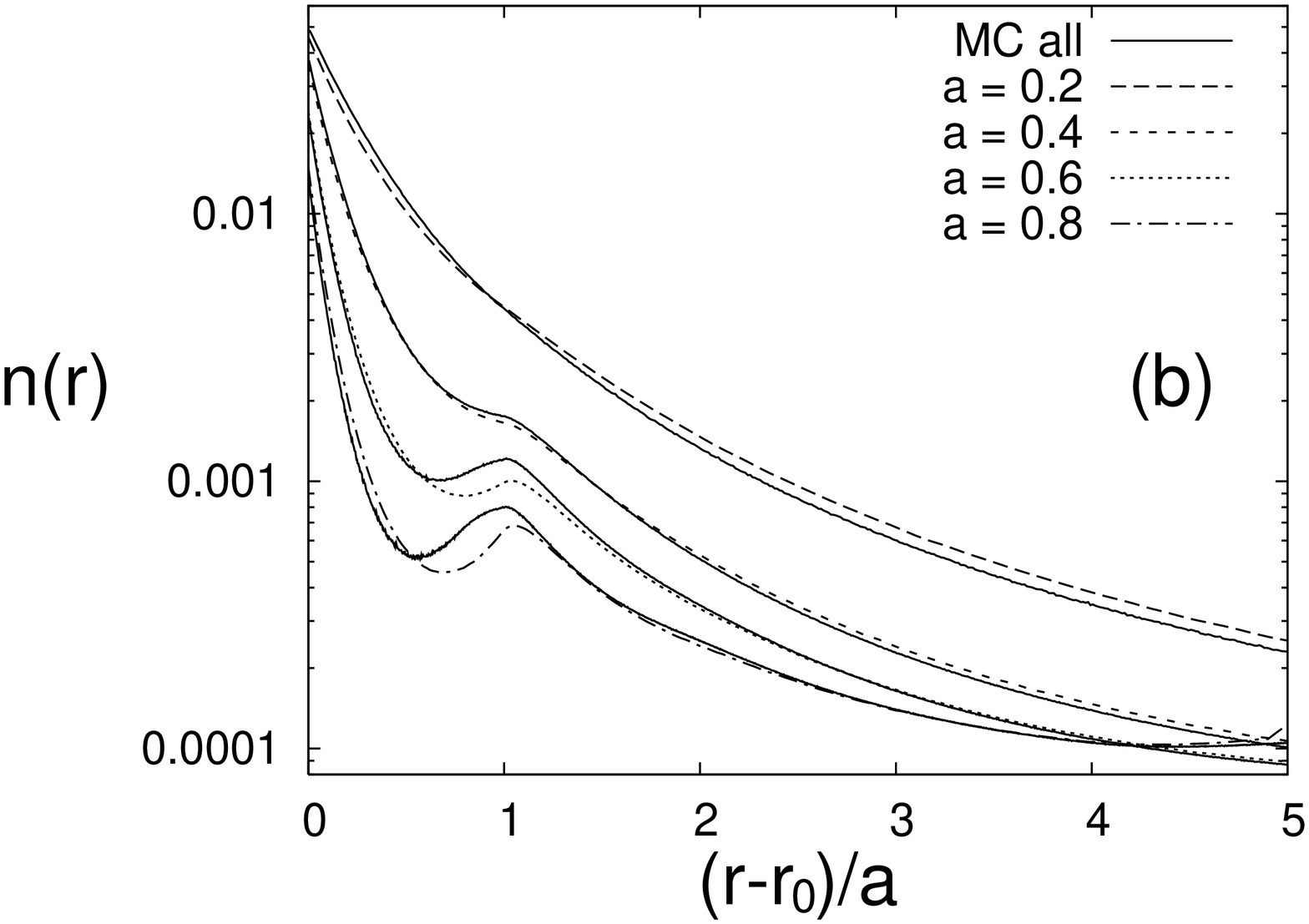}
  \caption{Ion distribution close to the colloid surface measured in systems with
   plasma parameter $\Gamma_{2d}=1.0$. Odd (a) and even (b) ionic diameters are shown
   separately for sake of clarity. Only WDA0 curves are marked while all MC curves
   are presented as solid lines and can be identified by the corresponding closest
   dashed WDA0 curve. All distances are measured in ionic diameters so the formation
   of the second layer of counterions is always expected to be around 1.}
  \label{n_r_lb2}
  \end{center}
  \end{figure}

Below we consider a model system in which parameters $r_0$ and
$R=5r_0$ are kept fixed, whereas the ion size and the Bjerrum
length are varied. The cell containing $N=100$ monovalent ions of
size ranging between $0.1r_0\leq a \leq 0.8r_0$ is studied at four
Bjerrum lengths $0.1r_0$, $0.2r_0$, $0.3r_0$ and $0.4r_0$. These
correspond to plasma parameter $0.5\leq \Gamma_{2d} \leq 2.0$
which enables us to investigate hard-core effects in systems with
weak, moderate and strong electrostatic correlations. The parameter
$\phi_s$ calculated for each of these 32 systems and given in
Table~\ref{tab} predicts that, at all Bjerrum lengths, the size
correlations are expected to be seen for ion sizes $a \geq
0.3r_0$, where $\phi_s$ is already greater than 0.2.
%
%
%
\begin{table}[tbp]
\begin{ruledtabular}
\begin{tabular}{|c||c|c|c|c|c|c|c|c|}
$l_B/r_0$ ($\Gamma_{2d}$)$\downarrow$ \; $a/r_0\rightarrow$ &  0.1
&  0.2 &  0.3 &  0.4 &  0.5 &  0.6 &  0.7 &  0.8 \\ \hline \hline
0.1 (0.5) & 0.01 & 0.08 & 0.23 & 0.44 & 0.74 & 1.13 & 1.59 &  2.13
\\ \hline 0.2 (1.0) & 0.02 & 0.11 & 0.28 & 0.53 & 0.87 & 1.29 &
1.79 &  2.37  \\ \hline 0.3 (1.5) & 0.03 & 0.13 & 0.31 & 0.57 &
0.92 & 1.35 & 1.86 &  2.46  \\ \hline
0.4 (2.0) & 0.03 & 0.13 & 0.32 & 0.59 & 0.95 & 1.38 & 1.91 &  2.51  \\
\end{tabular}
\caption{Parameter $\phi_s$ of different ionic sizes (columns),
Bjerrum (rows) lengths and plasma parameter (rows).
 Both Bjerrum lengths $l_B$
   and ion sizes $a$ are given in units of $r_0$. Packing effects are expected to be
   seen in those systems for which $\phi_s > 0.2$. Unrealistically high values of
   $\phi_s$ observed for large ions indicate inapplicability of PB and
   failure of LDA for these systems.}
\label{tab}
\end{ruledtabular}
\end{table}
%
%
The Monte Carlo data show that, indeed, the ion density profile is
concave for $a=0.1r_0$ and $a=0.2 r_0$ but develops a convex
region at distance about $a$ from the colloid surface for larger
ion sizes at all four Bjerrum lengths. This indicates some packing
taking place which is well captured by our $\phi_s$-criterion.
Figure~\ref{n_r_lb2}
%
%
  shows both the reference Monte Carlo and WDA0 density profiles calculated
  for different ion sizes at fixed Bjerrum length $l_B = 0.2r_0$
  ($\Gamma_{2d}=1.0$). The development of layering with increasing the ion
  size is well captured by the non-local functional approach, whereas none of the LDAs
  exhibits any layering and therefore are not shown in Fig.~\ref{n_r_lb2}.

\begin{figure}[!ht]
  \begin{center}
  \includegraphics[height=8.0cm,angle=-90]{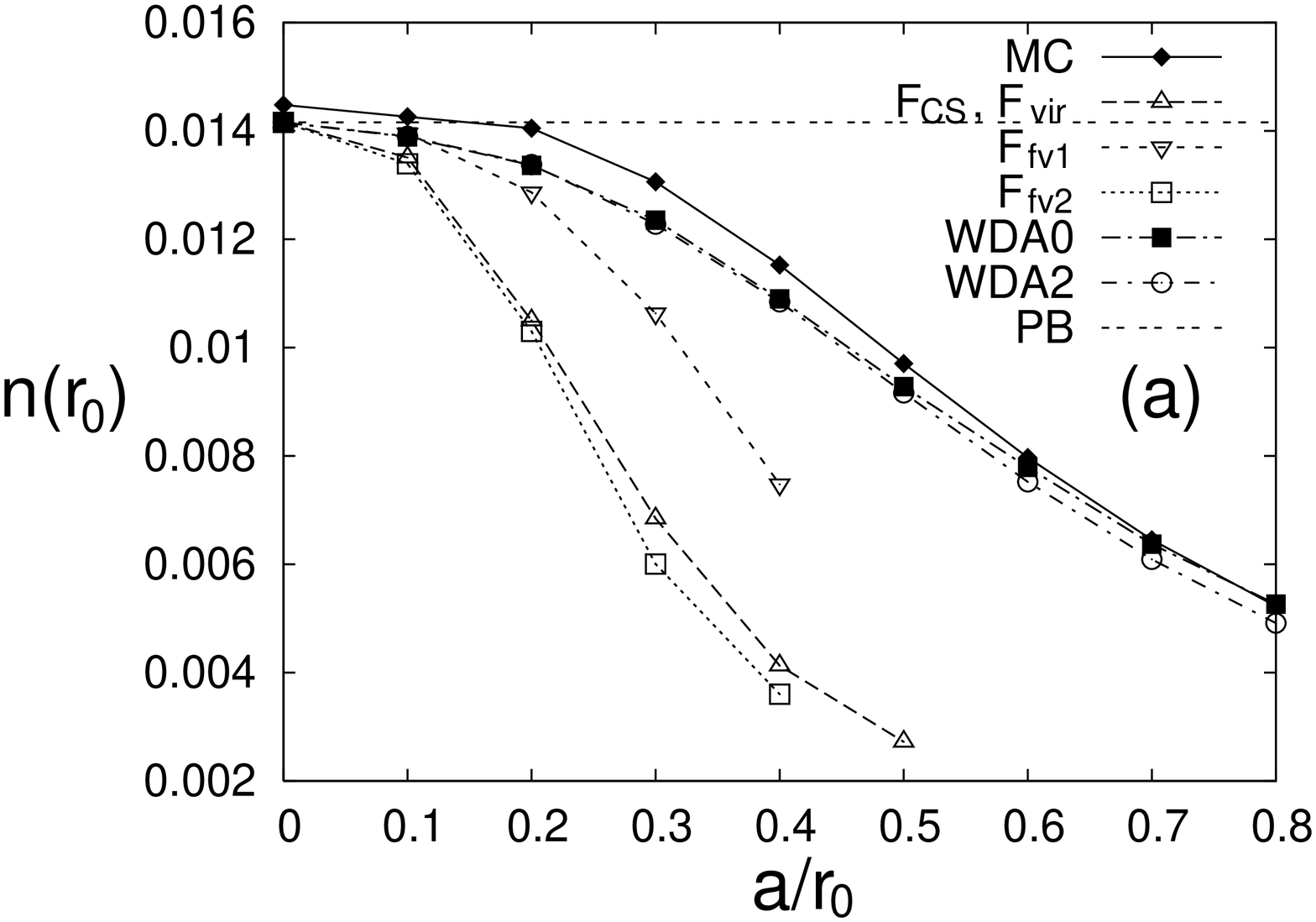}
  \includegraphics[height=8.0cm,angle=-90]{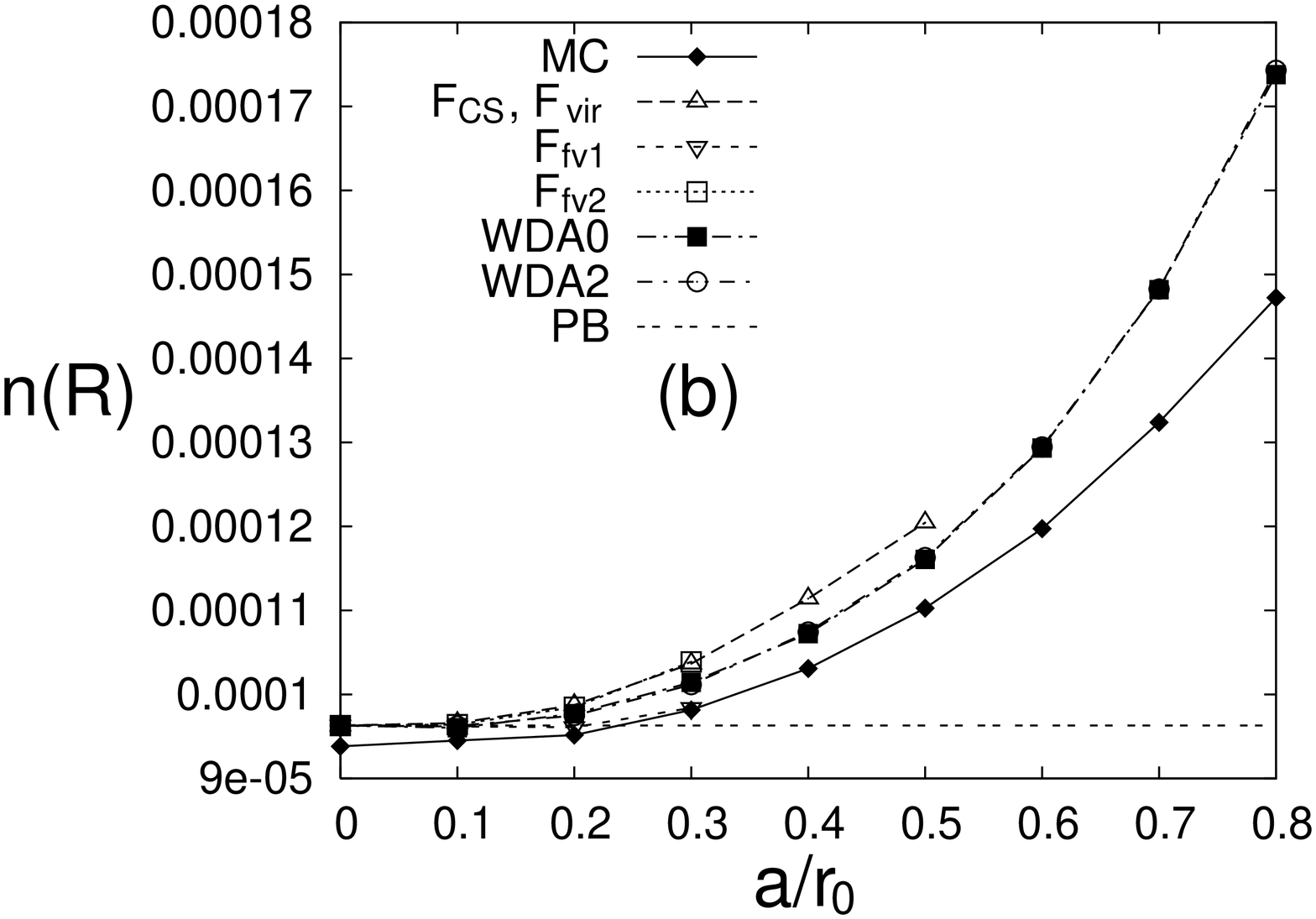}
  \caption{Contact density (a) $n(r_0)$ and boundary density (b) $n(R)$ predicted for systems
   with $\Gamma_{2d} = 0.5$ by different methods
   as a function of ionic diameter. All theories at $a=0$ are identical to PB which
   slightly underestimates the contact density due to ignoring electrostatic correlations.
   This effect reduces as the ion size increases. The legend is organized similar to that
   of Fig.~\ref{bor} with one more line WDA2 corresponding to the weight by Tarazona
   given by Eqs.~\ref{wtarazona}-\ref{wtarazona2}. There is a slight positive difference
   in $n(R)$ predicted by WDA2 and WDA0 not seen clearly in (b) at this scale.}
  \label{contact}
  \end{center}
  \end{figure}

Another way of checking how well correlations are captured by a particular
excess free energy functional is to compute the osmotic pressure. In real
systems this pressure also depends on correlations between ions of different
cells, something which is not taken into account within the cell model
approximation. So by pressure we refer to the pressure exerted on the rigid
wall at $r=R$ of our cell model. Within the simulations, the pressure is
given~\cite{wennerstroem82a}
by the contact density at $r=R$:
\begin{equation}
  \Pi = k_B T n(R) .
  \label{eq:pressure}
\end{equation}
For the density functional approach, this exact expression should
be corrected~\cite{tellez03a} to recover the free energy
functional after integrating pressure over the volume. The
correction term is generally small and for simplicity we will
directly compare contact densities predicted by different methods.
Figure~\ref{contact}
%
%
shows both the colloid contact density $n(r_0)$ and boundary density $n(R)$\
given by the Monte Carlo simulations and different local and non-local
density approaches for systems at fixed Bjerrum length $l_B =
0.1r_0$ ($\Gamma_{2d}=0.5$) as a function of the ionic diameter
$a$. The colloid contact density is informative of how well a
certain method works at the most packed region of the system and
can also be related to the pressure. Figure~\ref{contact}(b) shows
that the local approaches underestimate $n(r_0)$ at high ionic
radius when compared to the simulations. We found that even for
small ionic sizes PB density profile was closer to the MC reference
data than the results of any of the locally ``improved" functionals.
Moreover, the limitation
of the local approach is illustrated by the failure of the LDA's
to converge at large $a$ (no data are shown for large ions). Both
WDAs we used here give consistent results when compared to MC
simulations.

\begin{figure}[!ht]
  \begin{center}
    \includegraphics[height=8.0cm,angle=-90]{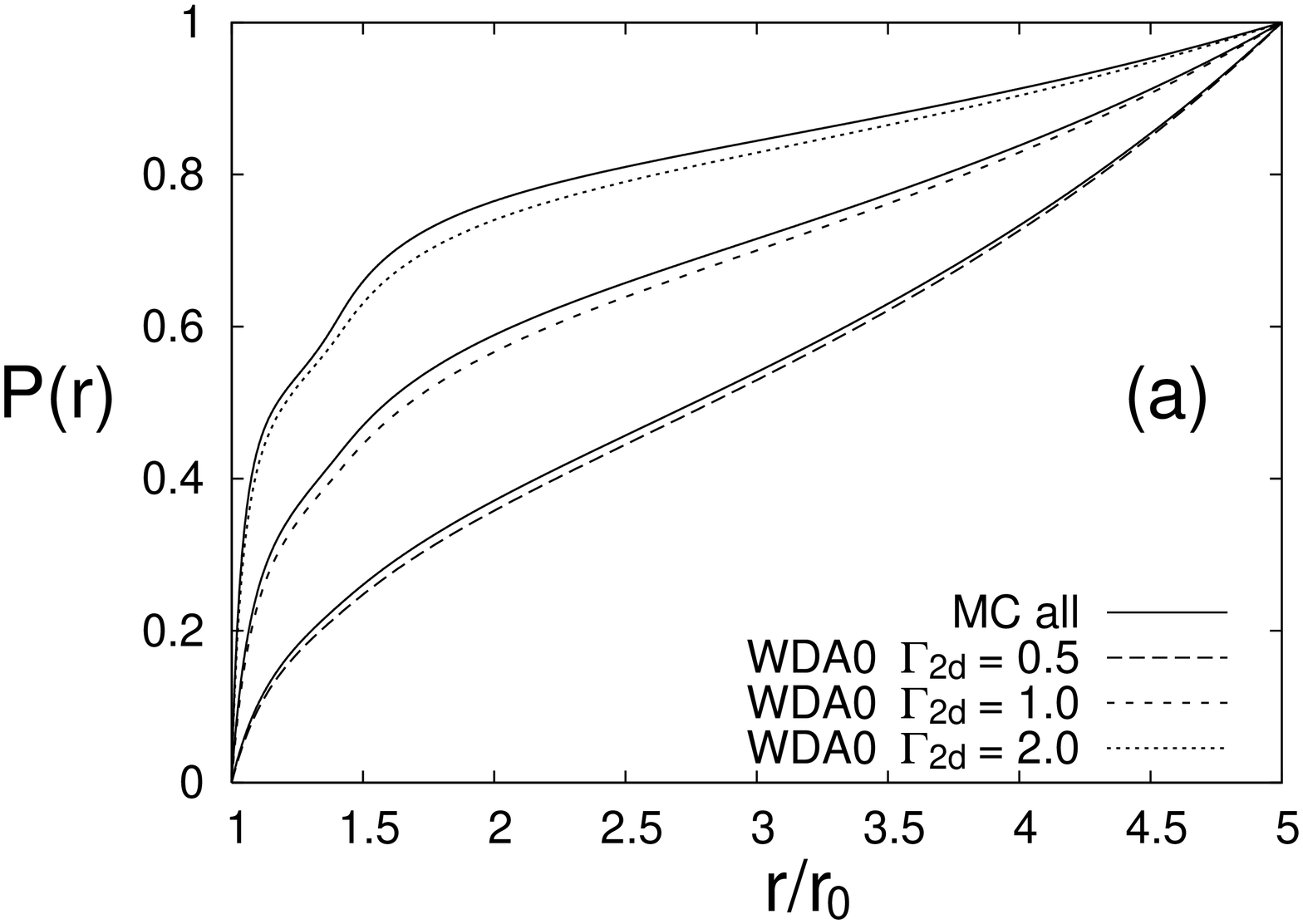}
    \includegraphics[height=8.0cm,angle=-90]{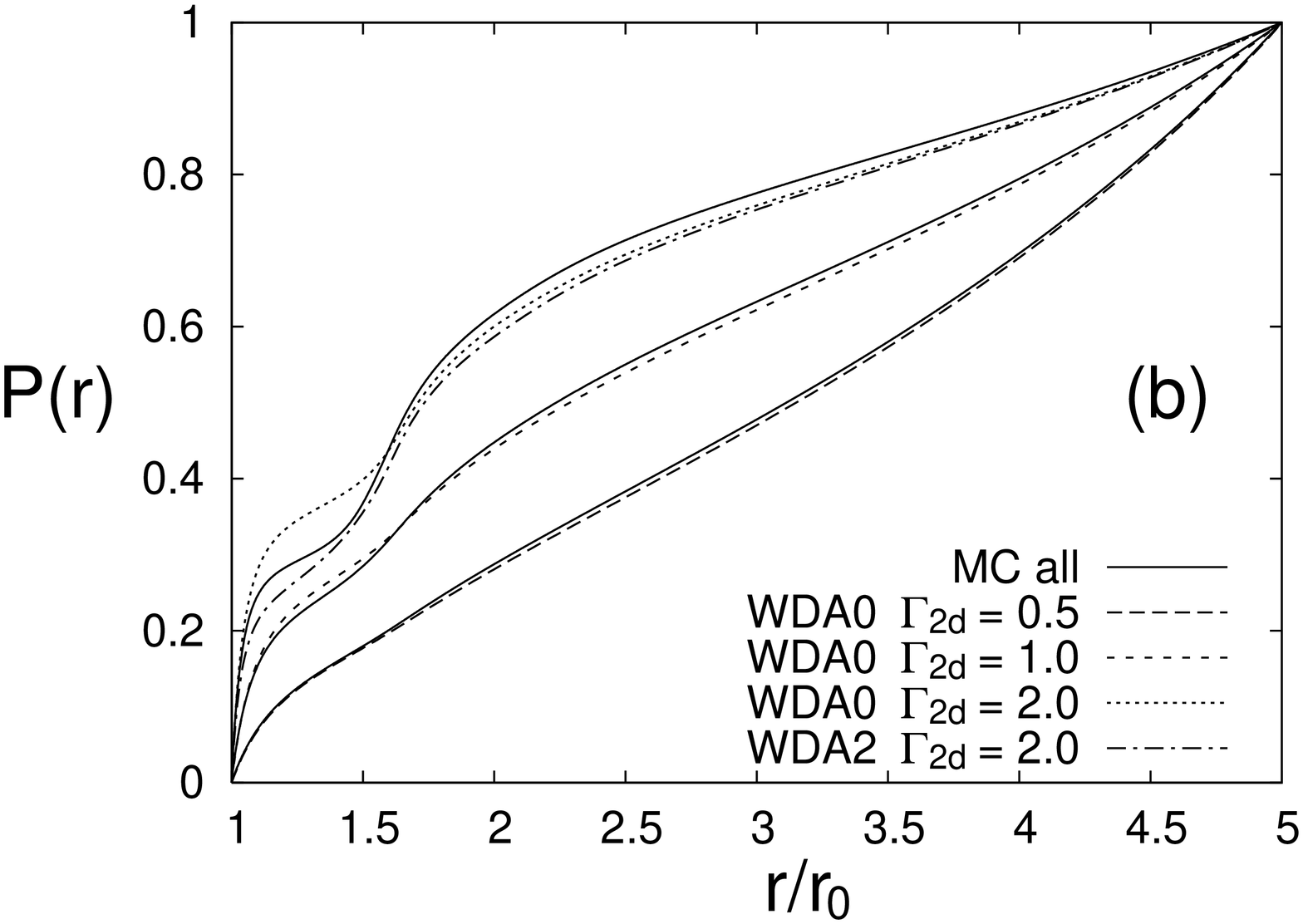}
  \caption{Integrated fraction of counterions obtained using Monte Carlo~(MC)
   and one constant weighted density approach~(WDA0) for plasma
   parameter $\Gamma_{2d}=0.5, 1.0, 2.0$ and for ionic diameters (a) $a=0.4r_0$
   and (b) $a=0.6r_0$. In (b), the WDA2 curve is also shown for the case of
   $\Gamma_{2d}=2.0$.}
  \label{P_r_a}
  \end{center}
\end{figure}

Now, we concentrate on the case of relatively large ions of
diameters $a \geq 0.4r_0$ for which layering is clearly observed.
According to the $\phi_s$-criterion, the density profiles of such
systems should deviate from PB and the hard-core interactions can
be even more significant than the electrostatics in some cases.
Figure~\ref{P_r_a} shows the integrated ion fraction
\begin{equation}
  P(r) = \frac{1}{N}\int_{r_0}^{r} dr \; 4\pi r^2 \, n(r) \;
  \label{eq:P}
\end{equation}
for systems with ionic diameters fixed at (a) $a=0.4r_0$ and (b)
$a=0.6r_0$ for plasma parameter $\Gamma_{2d}$ 0.5, 1.0 and 2.0.
%
%
%
%
Clearly, a larger plasma parameter leads to an increased condensation (the
curves are shifted up) -- an effect which is governed by the electrostatics
and also present in PB theory.  One could expect that for high $\Gamma_{2d}$
electrostatic correlations would be a dominant effect~\cite{barbosa04a}, and
significant deviations to PB theory and also to our WDA corrected DFT should
arise due to electrostatic correlations, which we did not account for.
However, under the investigated circumstances, the ionic size plays a more
relevant role.  The hard core effects lead to packing effects that
overcompensate the electrostatic correlations. For $a=0.4r_0$, the density
profiles are well captured by WDA0, which is always much closer to the MC data
than to the PB result (not shown here). However, for $a=0.6r_0$, the structure
of packing becomes important at $\Gamma_{2d} = 2.0$. For this system, the WDA2
weight improves the result, showing that the deviation between the simulations
and the constant weight WDA0 is not due to the electrostatic correlations but
rather to hard-core effects. Beyond this point a more sophisticated functional
should be used to capture the local packing.

\begin{figure}
  \begin{center}
    \includegraphics[height=9.0cm,angle=-90]{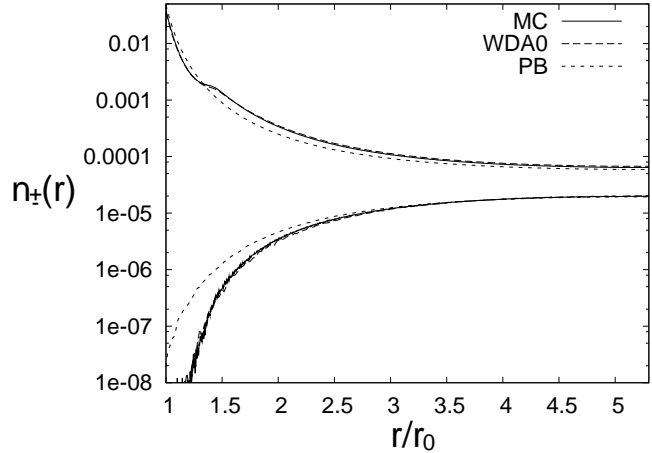}
  \caption{Density profiles of positive and negative ions obtained using MC simulations,
   WDA0 and PB for the system with $\Gamma_{2d} = 1.0$, $a = 0.4r_0$ and $10\%$ of salt.}
  \label{salt_G1}
  \end{center}
\end{figure}

\begin{figure}[!ht]
  \begin{center}
    \includegraphics[height=8.0cm,angle=-90]{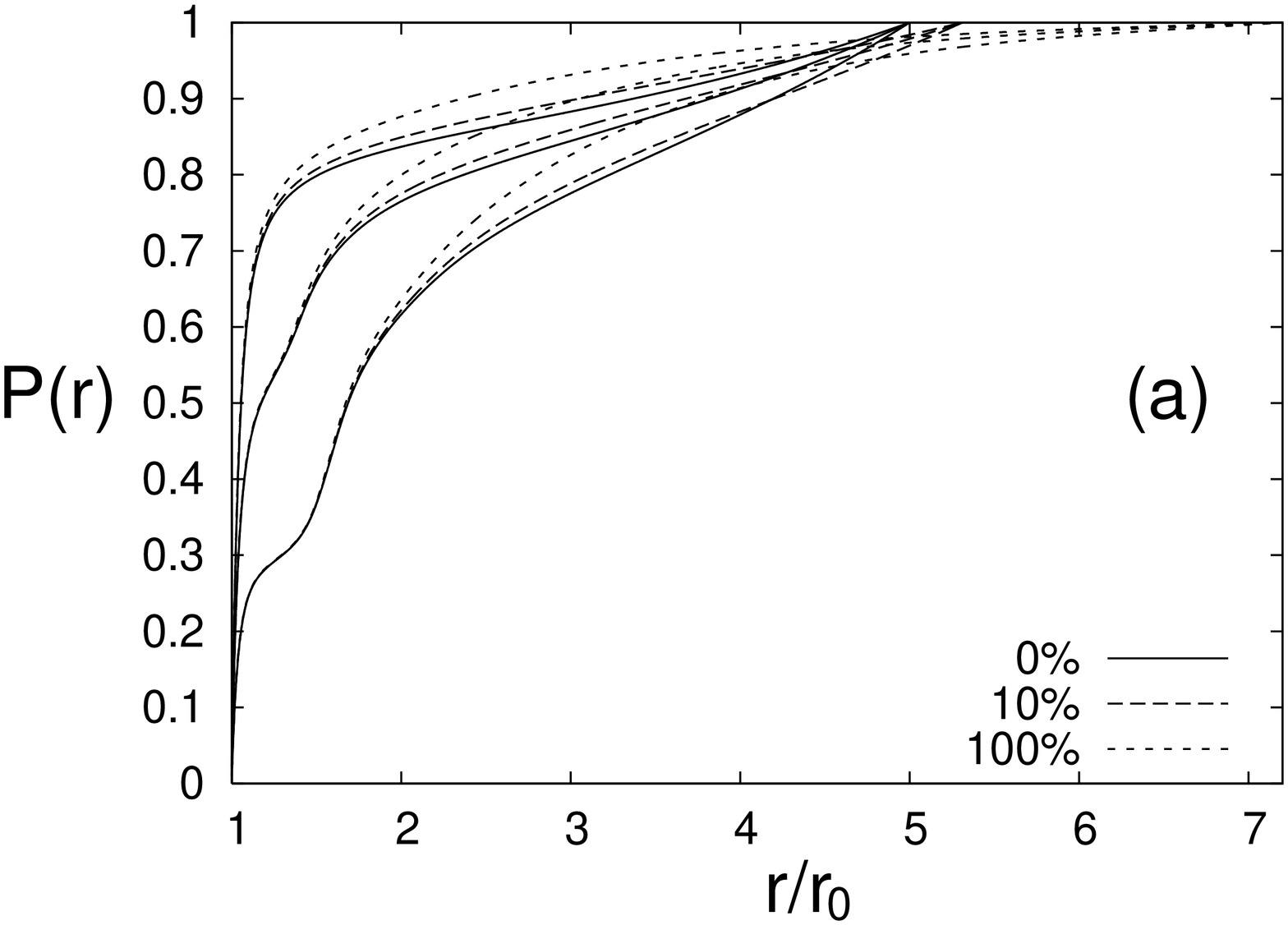}
    \includegraphics[height=8.0cm,angle=-90]{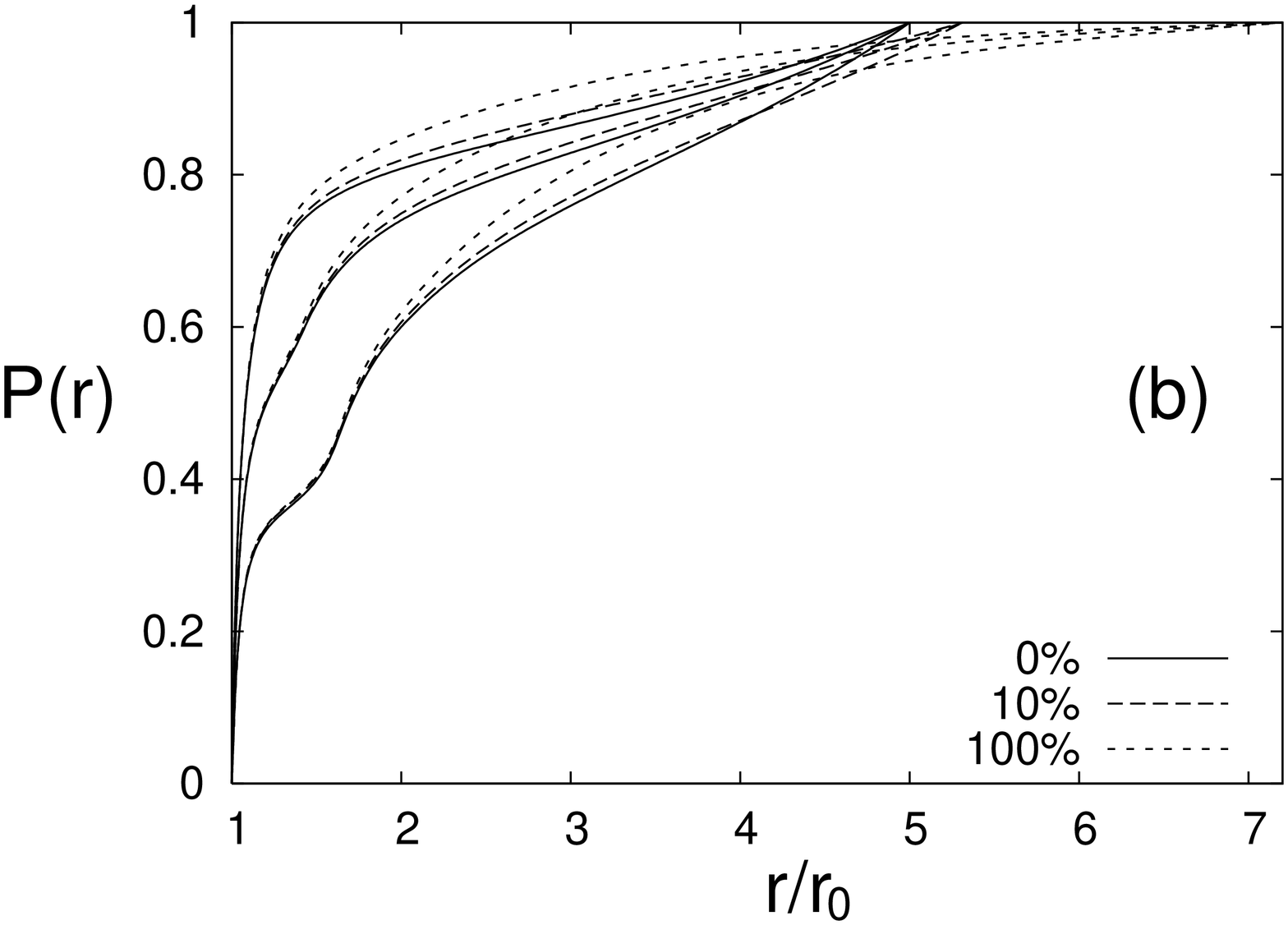}
  \caption{Integrated fraction of counterions obtained using (a) Monte Carlo~(MC)
   and (b) one constant weighted density approach~(WDA0) for different ion sizes
   at plasma parameter $\Gamma_{2d}=2.0$ and different amount of salt
  ($N_s$ = 0, $10\%$ and $100\%$ of $N$). The sizes vary from top to bottom as
   $a=0.1r_0$, $a=0.4r_0$ and $a=0.6r_0$,}
  \label{salt}
  \end{center}
\end{figure}

In principle, the addition of salt can lead to new correlations due to ion-ion
correlations and screening. For high electrostatic salt couplings, $\Gamma_s =
\frac{\ell}{d}$, where $d$ is the distance of closest approach of ion and
coion, ion clusters can also appear, that change the ion distribution
considerably~\cite{gonzalestovar85a,deserno01b}.  However, this effect can be
overcome by the hard core if the ions are large enough, rendering $\Gamma_s
\lesssim 1$. Below we consider the systems from Tab.~\ref{tab} with added
salt. Two cases of $N_s = 10$ ($10\%$ of salt) and $N_s = N$ ($100\%$ of salt)
are studied to represent moderate and high amount of salt. Since addition of
the salt ions into the cell would increase the packing fraction, here, we
prefer to keep it constant adjusting accordingly the cell radius $R$. This is
partially justified by the fact that the ion distribution close to the colloid
weakly depends on the cell size for our system parameters.
Figure~\ref{salt_G1} shows both the positive and negative charge density
profiles obtained using simulations, WDA0 and PB for the system with
$\Gamma_{2d} = 1.0$, $a = 0.4r_0$ with $10\%$ of salt.
%
%
%
%
Due to the screening, the effect of electrostatic correlations is less
profound than in the zero-salt case. The agreement between the simulations and
WDA is therefore improved at higher salt concentrations and lower plasma
parameter. The integrated charge profiles for several systems employing our
highest plasma parameter $\Gamma_{2d}=2.0$, are shown in Fig.~\ref{salt}.
%
%
%
%
The WDA captures the same trend in the charge distribution as provided by the
MC data. The presence of salt does not change the layering as is observed for
this ionic radius when the salt is not present. One should be aware that
things will be more complicated if one considers asymmetric salt (in valence
and size), or large ion sizes, since then more complicated effects like
overcharging can occur
\cite{gonzalestovar85a,greberg98a,deserno01b,messina02d}.

\section{Conclusion}
\label{conclusion}
In this paper we studied the effects of adding various local and non-local
free energy functionals to the PB free energy functional to include the
effects of an ionic hard core. We started from calculating a layer volume
fraction $\phi_s$ using the PB approximation, that gives a criterion for
recognizing when size effects become relevant and the simple PB approach has
to be modified. We tested this criterion for a system consisting of a charged
spherical colloid and its counterions confined to a spherical cell, and
studied a number of parameter combinations where the PB approximation fails.
For including size correlations, four local and two non-local density
functional approximations were employed. The local theories were always found
to overestimate the hard-core effects, creating an exclusion region close to
the colloid for large ionic radii. Beyond a certain ionic radius, all the
considered LDAs diverge and produce meaningless results. The failure of the
LDA is also captured by the increasing divergence between the LDA and the MC
contact densities which is seen when the ionic radius is increased, and the
absence of any layering effect in the LDA.  Due to this observations we note
that the inclusion of the LDA correction into PB actually~{\it worsens} the
agreement of PB with simulation results. On the other hand, we demonstrated
that a simple weighted density approach for the excluded volume interaction
was able to capture the main features of the ionic density profile. The
introductions of a more sophisticated weighted density approximations such as
Tarazona approach~\cite{tarazona85a} improves the agreement with the
simulation, but it does not bring any new physics to the problem. If some salt
is included, under certain parameters the main effect is the increase of
screening of the electrostatic correlations.  Therefore, the system can be
adequately described by the PB approach supplemented with an excluded volume
WDA. More complicated effects are expected to appear at sufficiently high
plasma parameters and higher salt concentrations. To treat those within
density functional theory, a combination of hard-core and electrostatic
correlations along the lines of Refs.~\cite{gillespie03a,yu02a} will probably
be required.

\section*{Acknowledgments}

This work has been supported by the Brazilian agencies, CNPq
(Conselho Nacional de Desenvolvimento Cient\'{\i}fico e
Tecnol\'ogico) and Capes through the process number 
210-05  and the German Science Foundation (DFG) through SFB
625, TR6 and Ho-1108/11-1, and PROBRAL contract PO-D-04-40434. We would like
to acknowledge M. Deserno for helpful discussions and MCB would also 
like to thank Prof.\ K.\ Kremer for the hospitality during her
stay in Mainz.

\appendix

\section*{Appendix}

The canonical partition function ${\cal Z}$ of the colloid
surrounded by its counterion in a cell model is given by:
\begin{equation}
  {\cal Z} =
  \int_{}^{} \prod_{i=1}^{N+N_s}  \prod_{j=1}^{N_s}\frac{d^{3}p_i d^3r_id^{3}p_j d^3r_j}{h^{3(N+N_s)}h^{3N_s}(N+N_s)!N_s!} \,e^{-\beta {\cal H}} \ ,
 \label{eq:ap1}
\end{equation}
where $N=Z/v$ is the total number of counterions, $2N_s$ is the
total number of positive and negative ions of salt. The
Hamiltonian ${\cal H}={\cal T}+{\cal V}$ splits into kinetic and
potential degrees of freedom. In the classical description
employed here the kinetic part ${\cal T}$ will contribute the
usual factor $\lambda^{-3N-6N_s}$ to the partition function, where
$\lambda$ is the thermal de Broglie wavelength. The potential
energy can be expressed as
\begin{eqnarray}
  {\cal V}
  & = &
  - N \sum_i^{N+N_s} \; \frac{\ell}{|{\bf{r}}_i|}
   + N \sum_i^{N_s} \; \frac{\ell}{|{\bf{r}}_i|}
  + \frac{1}{2} \sum_{i\ne j}^{N_s,N_s}\; \frac{\ell}{|{\bf{r}}_i-{\bf{r}}_j|} \\
\nonumber
 &+& \frac{1}{2} \sum_{i\ne j}^{N_s+N,N_s+N} \; \frac{\ell}{|{\bf{r}}_i-{\bf{r}}_j|}
-\frac{1}{2} \sum_{i\ne j}^{N_s,N_s+N} \; \left[
\frac{\ell}{|{\bf{r}}_i-{\bf{r}}_j|} +g(|{\bf{r}}_i-{\bf{r}}_j|/a) \right]
\;,
 \label{eq:ap4}
\end{eqnarray}
where the first two terms are related to the electrostatic
interactions and the last is responsible for the hard-core
repulsion. The specific form of this term is not relevant here. 
After rescaling all length by $\ell$, i.e.
introducing ${\bf{\hat{r}}} := {\bf{r}}/\ell$, the total partition
function can be rewritten as
\begin{eqnarray}
  {\cal Z} & = &
  \frac{1}{(N+N_s)!N_s!}\left(\frac{\ell}{\lambda}\right)^{3N+6N_s}
  \int_{\hat{r}_0}^{\hat{r}_0/\phi^{1/3}} \prod_k^{N+N_s}  \prod_l^{N_s}\; \int_{}^{}d^3 \hat{r}_kd^3 \hat{r}_l
  \\ \nonumber
  & & \exp\bigg\{
  - N \sum_i^{N+N_s} \; \frac{1}{|{\bf\hat{r}}_i|} + N \sum_i^{N_s} \; \frac{1}{|{\bf\hat{r}}_i|}+\frac{1}{2} \sum_{i\ne j}^{N_s+N,N_s+N} \; \frac{1}{|{\bf\hat{r}}_i-{\bf\hat{r}}_j|}\\ \nonumber
  &+&\frac{1}{2} \sum_{i\ne j}^{N_s,N_s} \; \frac{1}{|{\bf\hat{r}}_i-{\bf\hat{r}}_j|}
-\frac{1}{2} \sum_{i\ne j}^{N_s+N,N_s} \;\left[
\frac{1}{|{\bf\hat{r}}_i-{\bf\hat{r}}_j|} +
g(|{\bf\hat{r}}_i-{\bf\hat{r}}_j|/\hat{a})\right]\bigg\} \;,
  \nonumber
\end{eqnarray}
where $\hat{a}=a/\ell$.

In this form it becomes evident that appropriately scaled thermal
observables like the integrated charge density (measured in units
of $\ell^{-3}$) or the pressure (measured in units of $k_B
T\ell^{-3}$) are invariant under system changes which keep the
number of counterions $N$, the number of salt particles $N_s$, the
rescaled colloid size $\hat{r}_0=r_0/\ell$, the rescaled ion
radius $\hat{a}$, and the volume fraction $\phi$ constant.

Poisson-Boltzmann theory shows the same invariance property, as
does the approximate density functional theory we are proposing in
this paper.


\bibliography{../bibtex/polyelectrolyte}


\end{document}